\documentstyle[prb,aps,epsf]{revtex}

\begin{document}
\draft
\title{A nature of low-temperature resistivity minimum in ceramic manganites}
\author{E. Rozenberg${}^{1}$, M. Auslender${}^{2}$, I. Felner${}^{3}$ and G.
Gorodetsky${}^{1}$}
\address{${}^{1}$Department of Physics Ben-Gurion University of the Negev\\
POB 653, 84105 Beer-Sheva, Israel}
\address{${}^{2}$Department of Electrical and Computer Engineering Ben-Gurion\\
University of the Negev\\
POB 653, Beer-Sheva 84105, Israel}
\address{${}^{3}$Racah Institute of Physics The Hebrew University\\
Jerusalem 91904, Israel }
\date{today}
\maketitle

\begin{abstract}
Measurements of magnetoresistance and magnetization were carried out on
ceramic samples of La$_{0.5}$Pb$_{0.5}$MnO$_{3}$ and La$_{0.5}$Pb$_{0.5}$MnO$%
_{3}$, containing 10 at. \% Ag in a dispersed form. The results obtained for
the resistivity at zero applied magnetic field exhibit a shallow minimum at
the temperature $T\approx 25\div 30$ K which shifts towards lower
temperatures upon applying a magnetic field and disappears at a certain
field $H_{{\rm cr}}$. Also the resistivity at helium temperature decreases
upon applying magnetic fields. It is shown that the model of charge carriers
tunneling between antiferromagnetically coupled grains may account for the
results observed.
\end{abstract}


\section{Introduction}

As with any material, the transport properties of doped manganites La$_{1-x}$%
A$_{x}$MnO$_{3}$ (A is a divalent ion such as Ca, Sr, Ba etc.) differ
markedly whether they are in single-crystalline or polycrystalline form (see
Chapters 1 and 5 in Ref.\cite{1}). There exists experimental evidence that
the presence of grains and grain boundaries (GB) modifies drastically the
type of the temperature and magnetic field dependence of the resistivity $%
\rho (T,H)$ in ceramic manganites as compared to single-crystalline samples.%
\cite{1,2,3} In single crystals (at $0.2<x<0.5$) a strong peak of $\rho
(T,0) $ is observed near the Curie temperature $T_{{\rm c}}$ - a
metal-insulator transition occurs. In contrast, the resistivity of a
polycrystalline samples exhibits a wide maximum at a temperature $T_{{\rm %
\max }}$ well below $T_{{\rm c}}$. At sufficiently small grain size no peak
is seen near $T_{{\rm c}} $, but as the grain size increases the peaks at
both $T_{{\rm \max }}$ and $T_{{\rm c}}$ become coexistent. \cite{4,5} In
good polycrystalline samples the former peak at $T_{{\rm \max }}$
degenerates to a ''shoulder'' at $T<T_{{\rm c}}$ and only a sharp maximum of 
$\rho (T,0)$ is observed near $T_{{\rm c}}$.\cite{4,5,6} The
magnetoresistance (MR) in the single crystals and polycrystalline samples
also behave very differently. E.g., the single crystals have colossal MR
(CMR) in a vicinity of $T_{{\rm c}}$ and small enough MR apart it.\cite
{1,2,3} On the other hand, the ceramics have an appreciable MR throughout
the ferromagnetic region and often manifest the largest MR at low
temperatures.\cite{1,4,7,8,9} This low-temperature MR is characterized by
enhanced low-field response $\Delta \rho /\Delta H$.\cite{1,7,8} In
addition to such low-field low-temperature MR, the prominent feature that
distinguish the polycrystalline from single-crystalline 
La$_{1-x}$A$_{x}$MnO$_{3}$ is a shallow minimum of $\rho (T,0)$, which occurs 
at a temperature $T_{\min }$ well below $T_{{\rm \max }}$. \cite{6,9,10,11,12}

Two approaches have been used to explain the phenomena specific for
polycrystalline manganites. Intergrain tunneling concept has been applied to
model the low- and high-field MR at low $T$,\cite{14} as well as the
temperature and grain-size dependencies of the resistivity.\cite{4,5} MR at
intermediate $T$ (around $T_{\max }$) and the low-temperature minimum of 
$\rho (T,0)$ have not been examined within this model. The latter phenomenon,
however, has been considered using the second approach, namely,
bulk-scattering concept.\cite{11,12} It has been suggested that the minimum
(and the resistivity upturn at lower $T$) arises from the competition of two
contributions - one, usual, increasing and other, decreasing with the
increase of the temperature. The origin of such an unusual contribution has
been attributed to Coulomb interaction (CI) between carriers strongly
enhanced by disorder.\cite{15,16} However, to the best of our knowledge, no
attempt was done to describe within CI model the flattening and vanishing of
the minimum under rather small external magnetic fields observed in very
different polycrystalline manganites.\cite{6,8,10}

The present paper focuses on the study of the low-temperature minimum of
resistivity in ceramic manganites. In Section 2 we present the experimental
results on the temperature dependence of the resistivity in zero and an
increased magnetic field. The data obtained are analyzed in Section 3 using
a bulk-scattering model ({\bf A}) and a model of carrier tunneling between
antiferromagnetically (AFM) coupled ferromagnetic (FM) grains ({\bf B}). The
results of the consideration are concluded in Section 4.

\section{EXPERIMENTAL RESULTS}

Samples of La$_{0.5}$Pb$_{0.5}$MnO$_{3}$ (LPMO) and LPMO containing 10 at.
\% of dispersed Ag were prepared \cite{13} by a standard ceramic technology.
It was noted in \cite{13} that Ag doping leads to the formation of Ag
agglomerates within the sample that, in turn, decrease the resultant
resistivity of LPMO (see Fig.1) but have no pronounced effect on the magnetic
and MR properties. Measurements of $\rho $ vs. T at zero magnetic field and
at $H$ up to $1.5$ Tesla were carried out in the temperature range $4.2-360$
K, and the results obtained are presented in Figs. 1 and 2. 
\vskip .5cm
\begin{figure}
\epsfxsize=3truein
\centerline{\epsffile{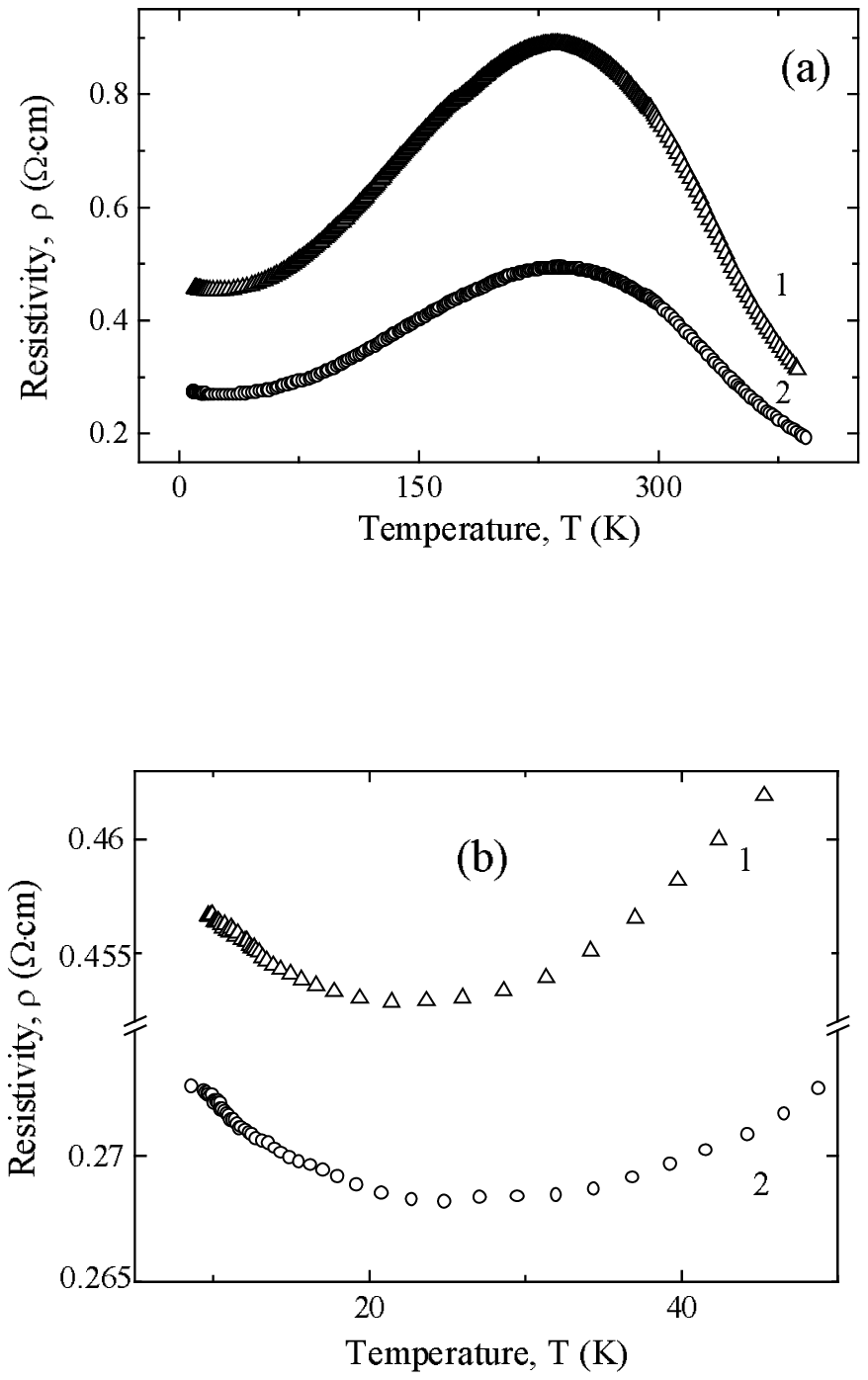}}
\label{Fig.1}
\end{figure}
%\vskip -11cm
\noindent
{\footnotesize {\bf FIG. 1.} 
(a) temperature dependence of the resistivity of La$_{0.5}$Pb$_{0.5}$MnO$_{3}$ 
(curve 1) and La$_{0.5}$Pb$_{0.5}$MnO$_{3}$ containing 10
at. \% of Ag in a dispersed form (curve 2); (b) - the temperature dependence
at low-temperature extended scale} 
\vskip .5cm

As it is shown in Fig. 1b the minimum of $\rho (T,H)$ is observed at 
$T_{\min }\sim 25\div 30$ K. This result is similar to that obtained on self-doped 
(with cation vacancies on La and Mn-sites),\cite{8} Ca-,\cite{10,12} Sr- \cite{6,12} 
and Ba-doped \cite{12} ceramic manganites, as well as on Ca-doped
polycrystalline films.\cite{9} Moreover, artificially created single grain
boundary induces the appearance of similar minimum for an epitaxial
bicrystal La$_{0.67}$Ca$_{0.33}$MnO$_{3}$ film.\cite{17} It seems that the
existence of the above low-temperature minimum is not sensitive to the
nature of La-site dopant and to the presence of additional impurities; its
position is almost sample-independent and is close to the above mentioned
values of $T_{\min }$. High level La-site doping of the LPMO ceramic do not
suppress the above noted wide maximum of $\rho (T,0)$ as well (see Fig.1).
\vskip .5cm
\begin{figure}
\epsfxsize=3truein
\centerline{\epsffile{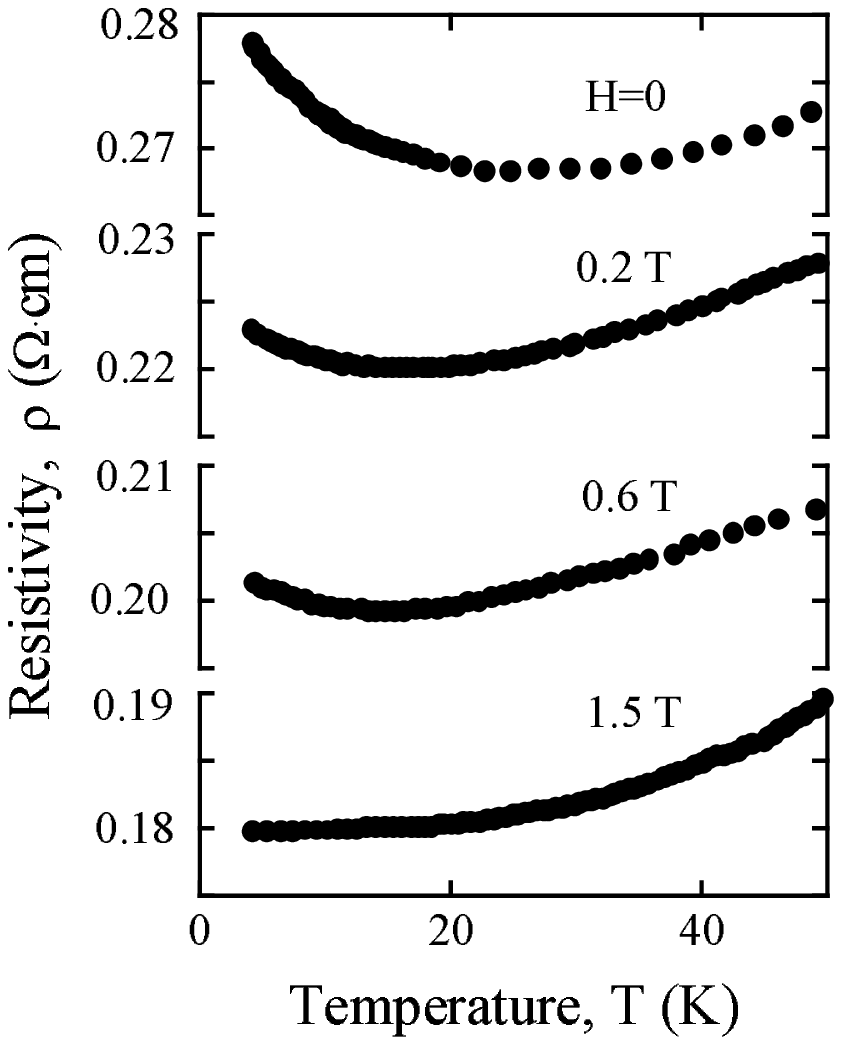}}
\label{Fig.2}
\end{figure}
%\vskip -11cm
\noindent
{\footnotesize {\bf FIG. 2.} 
Experimental temperature dependence of the resistivity of 
La$_{0.5}$Pb$_{0.5}$MnO$_{3}$ containing 10 at. \% of Ag measured at various
magnetic fields} 
\vskip .5cm

The measurements of the magnetization $M$ vs. $T$ at a low $H$ were reported
previously. \cite{13} For the benefit of the present discussion, the
low-temperature data are displayed here in an extended scale, see Fig. 3.
\vskip .5cm
\begin{figure}
\epsfxsize=3truein
\centerline{\epsffile{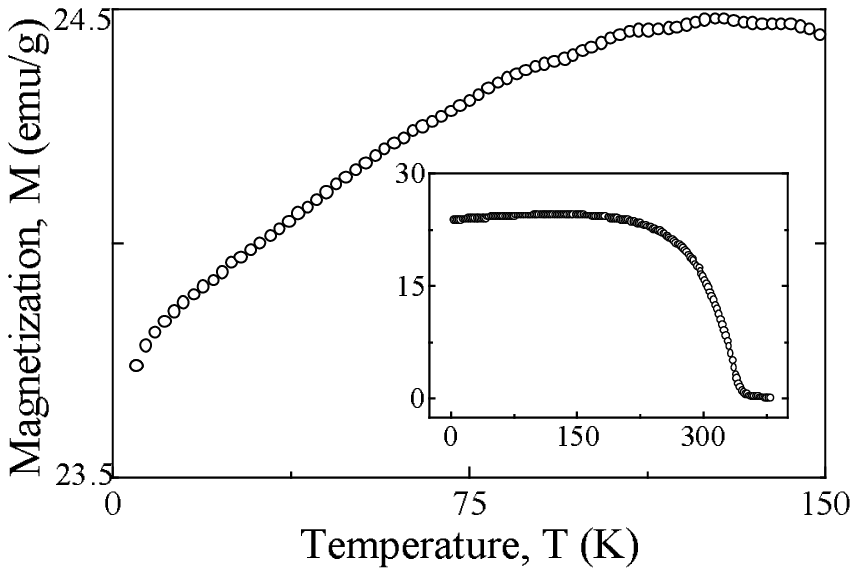}}
\label{Fig.3}
\end{figure}
%\vskip -11cm
\noindent
{\footnotesize {\bf FIG. 3.} 
Magnetization vs. temperature measured at low magnetic field.
The results are presented at temperatures up to 150 K; the magnetization vs.
temperature measured at a wider temperature region is given in the inset}
\vskip .5cm

The MR and magnetic properties of the LPMO ceramic\cite{13} are similar to
those of other ceramic manganites.\cite{4,5} In particular, $T_{\max }$ is
significantly lower than $T_{{\rm c}}$, the largest MR effect occurs at
lowest $T$, and, except only a small peak near $T_{{\rm c}}$, MR gradually
decreases with increasing temperature. Note that the resistivity upturn at 
$T<T_{\min }$ could not be attributed to the effect of charge ordering
since, similarly to that observed in other ceramic manganites, \cite{6,10,12}
it appears to be weak ($<1\%$). So, the LPMO is an appropriate system for
studying effects, which are peculiar for the La-based manganite ceramics.

\section{ANALYSIS OF THE RESULTS}

\subsection{Bulk-scattering model including quantum corrections to
conductivity}

Let us first analyze $\rho (T,H)$ at low temperatures under an assumption
that bulk scattering dominates the conduction. It seems reasonable in the
ceramics with large contacting grains that may form a percolation cluster.
Generally, the resistivity can be represented as 
\begin{equation}
\rho =\rho _{{\rm el}}+\rho _{{\rm in}}  \label{resist}
\end{equation}
where $\rho _{{\rm el}}$ and $\rho _{{\rm in}}$ are the contributions due
elastic (electron-impurity interaction and CI) and inelastic (e.g.
electron-phonon interaction) processes. Normally, the part $\rho _{{\rm in}}$
increases with increasing the temperature due to a power law $\rho _{{\rm in}%
}=bT^{p}$ in which the coefficient $b$ doesnt depend on $H$. In good
conductors the part $\rho _{{\rm el}}=1/\sigma _{{\rm el}}$ doesn't depend
on $T$ and $H$ being equal to the residual resistivity $\rho _{0}$, but at
rather strong disorder both temperature and magnetic-field dependence
appears in $\sigma _{{\rm el}}$ due to CI and decoherence effects.\cite
{15,16}

A theory of CI correction to the residual conductivity at $H=0$ was
originally developed by Altshuler and Aronov \cite{15}, yielding the
following expression 
\begin{equation}
\sigma _{{\rm el}}\left( T,0\right) =\sigma _{0}+\delta \sigma _{{\rm CI}%
}\left( T,0\right) ,\quad \delta \sigma _{{\rm CI}}\left( T,0\right) =0.0309%
\frac{e^{2}}{\hbar L_{{\rm T}}}  \label{cicorr}
\end{equation}
where $L_{{\rm T}}^{-1}=\sqrt{k_{{\rm B}}T/D}=A\sqrt{T}$ and $D$ is the
carriers diffusion constant. Thus, with the use of Eq.(\ref{cicorr}) and
assumed form of $\rho _{{\rm in}}$ , Eq.(\ref{resist}) takes the form 
\begin{equation}
\rho \left( T,0\right) =\rho _{0}-aT^{1/2}+bT^{p}  \label{reswimin}
\end{equation}
where $a=0.0309A\rho _{0}^{2}e^{2}\hbar ^{-1}$. The interplay of the
increasing and decreasing temperature-dependent terms in Eq.(\ref{reswimin})
gives rise to the minimum of $\rho (T,0)$. Such an explanation was suggested
in Refs.\cite{11,12} It appears that our experimental curves $\rho (T,0)$ at
low temperatures are fitted fairly well by Eq.(\ref{reswimin}) with the same 
$p=2$ as was taken in Refs.\cite{11,12} Though, this explanation is
plausible, it must be verified by an additional evidence. In this connection
the influence of external magnetic field on the minimum is a crucial test
for the model. Very recently, the calculation of the total quantum
correction to $\sigma _{0}$ that could shed light on our problem, has been
accurately recast by Aleiner et al.\cite{18} It gives 
\begin{equation}
\sigma _{{\rm el}}\left( T,H\right) =\sigma _{0}+\delta \sigma _{{\rm CI}%
}\left( T,H\right) +\delta \sigma _{{\rm DP}}\left( T,H\right) 
\label{signzh}
\end{equation}
where $\delta \sigma _{{\rm CI}}\left( T,H\right) $ is the CI contribution
in a non-zero magnetic field, and $\delta \sigma _{{\rm DP}}\left(
T,H\right) $ is the contribution of dephasing caused by both magnetic field
and CI. Asymptotic formulas of Ref. \cite{18}, suitable mostly for 1D and 2D
cases, contain the terms divergent in 3D. To get finite results such terms
have been recalculated. For the CI correction we have 
\begin{equation}
\frac{\delta \sigma _{{\rm CI}}\left( T,H\right) }{\delta \sigma _{{\rm CI}%
}\left( T,0\right) }-1=\frac{0.7510}{k_{{\rm F}}^{2}l\,l_{{\rm H}}}
\label{ciresnozh}
\end{equation}
where $k_{{\rm F}}$ is the Fermi wave-number, $l$ is the mean free path and $%
l_{{\rm H}}=\sqrt{c\hbar /eH}$ is the magnetic length. The condition $l_{%
{\rm H}}>>$ $L_{{\rm T}}$ under which the calculation was done\cite{18}
holds for $H$ up to tens Tesla. Formally Eq.(\ref{ciresnozh}) predicts a
reduction of the CI correction but actually this effect is negligibly small.
For example, at typical carrier density  $\sim 10^{22}$ cm$^{-3}$, $k_{{\rm F%
}}l>1$, the right-hand side of Eq.(\ref{ciresnozh}) is smaller than $1\%$ up
to $H=10$ T. The term $\delta \sigma _{{\rm DP}}\left( T,H\right) $ has been
obtained assuming domination of the magnetic-field contribution to this
correction.\cite{18} The result may be treated as first two terms of
expansion of the weak-localization expression 
\begin{equation}
\delta \sigma _{{\rm DP}}\left( T,H\right) =\frac{e^{2}}{\left( 2\pi \right)
^{2}\hbar }\sqrt{4l_{{\rm H}}^{-2}+L_{\varphi }^{2}}  \label{dephasht}
\end{equation}
with respect to small $l_{{\rm H}}/L_{\varphi }$,\cite{15,16,18} where $%
L_{\varphi }$ is the length of dephasing due to CI given by 
\begin{equation}
L_{\varphi }^{-2}=\frac{14.7336}{k_{{\rm F}}^{2}l\,L_{{\rm T}}^{3}}\left(
1-1.2794\frac{2L_{{\rm T}}}{l_{{\rm H}}}\ln \frac{l_{{\rm H}}}{2L_{{\rm T}}}%
\right)   \label{dephlength}
\end{equation}
The numerical factors in Eqs.(\ref{cicorr}), (\ref{ciresnozh}) and (\ref
{dephasht}) stem from the assumed parabolic spectrum and full spin
polarization of the carriers. Eq.(\ref{dephasht}) may serve as a reliable
extrapolation of $\delta \sigma _{{\rm DP}}\left( T,H\right) $ to all $H$.
It follows from Eq.(\ref{dephasht}) that parametrically $L_{{\rm T}}$ $%
<<L_{\varphi }$, and so $\delta \sigma _{{\rm DP}}\left( T,H\right) <<\delta
\sigma _{{\rm CI}}\left( T,H\right) $ (for realistic parameters these
inequalities may not hold in the strong form).

So, in the considered model at all actual $H$ the resistivity minimum
persists and is weakly affected by the magnetic field. It is worth to note
that such a resistivity minimum is observed in a single-crystalline
bilayered manganite.\cite{19} At the same time the behavior noted above
fully disagrees with the experimental data for ceramic manganites (see
Sections 1 and 2). Magnetic-field dependence of $\delta \sigma _{{\rm DP}%
}\left( T,H\right) $ leads to a negative MR, but it is too small to account
for the observed MR. At $l_{{\rm H}}<<L_{\varphi }$ $\,\delta \sigma _{{\rm %
DP}}\left( T,H\right) /\sigma _{0}\approx 3(k_{{\rm F}}^{2}l\,l_{{\rm H}%
})^{-1}$ and hence, as follows from the above estimation, the model MR in
rather strong fields achieves percents at best. In comparison, measured MR
is  $\sim 36\%$ at $T=4.2$ K, and $H$ of only $1.5$ T causes the minimum to
disappear (see Fig. 2).

Thus, the bulk-scattering model with quantum corrections to conductivity
predicting a resistivity minimum at zero magnetic field, strongly disagrees
with the experiment on ceramic manganites as concerns the behavior of the
minimum in finite magnetic fields.

\subsection{Intergrain tunneling model}

Let us treat the problem using an approach, which is quite opposite to that
examined previously. Namely, we will assume that in the discussed materials
much of the grains are isolated from each other. In this model the tunneling
between grains brings dominating contribution to the conduction. This is
consistent with the low-temperature conductivity weakly dependent on T, but
being smaller than the minimal metallic conductivity.\cite{6} Several groups
(see e.g. Ref.\cite{1}) attributed the low-temperature MR to spin-polarized
intergrain tunneling. Our purpose is to describe the low-temperature
resistivity minimum also using this concept.

Regular theory of tunneling conduction through FM metal - non-magnetic
insulator - FM metal junction exists only for the planar geometry. \cite
{20,21} For the tunneling resistance between two FM grains, say $i$ and $j$,
we will use phenomenological expression of Ref.\cite{7} 
\begin{equation}
R_{ij}=\frac{r_{ij}}{1+\varepsilon \cos \vartheta _{ij}}.  \label{tunresis}
\end{equation}

Here $\vartheta _{ij}$ is the angle between the magnetization directions $%
{\bf n}_{i}$ and ${\bf n}_{j}$ of the grains $i$ and $j$, respectively ($%
{\bf n}_{i}^{2}=$ ${\bf n}_{j}^{2}=1$), $\varepsilon $ = $P^{2}$ is the
spin-valve coefficient, \cite{20,21} where $P$ is the degree of spin
polarization of current carriers in each grain 
\begin{equation}
P=\frac{N_{\uparrow }-N_{\downarrow }}{N_{\uparrow }+N_{\downarrow }},
\label{spinpol}
\end{equation}
$N_{\uparrow }$ and $N_{\downarrow }$ being the densities of states for
spin-up and spin-down carriers, respectively, and $r_{ij}$ is a factor
independent on the magnetization orientations, which cannot be determined in
phenomenological approach. By analogy with the case of planar-junction
tunneling \cite{20,21} it has been proposed \cite{4,5} to use the expression 
\begin{equation}
r_{ij}=r_{0}\exp \left( 2\kappa d_{ij}\right) ,\;\kappa =\sqrt{2mU}/\hbar
\label{rfactor}
\end{equation}
where $d_{ij}$ is the distance between the grains and $U$ is the tunneling
barrier height. Though Eq.(\ref{rfactor}) holds only at large $d_{ij}$
and/or $U$, it may be used for qualitative description.

Eqs.(\ref{tunresis}) and (\ref{rfactor}) define a random resistor network,
which has to be solved in order to obtain the sample resistivity. As it
stands, the problem cannot be solved unless the statistics of the
configurations $\{{\bf n}_{i}\}$ is defined. At equilibrium the
configuration energy $E$ and temperature $T$ defines the statistical
distribution. One of the contributions to $E$ is the energy of the grain
moments (${\bf m}_{i}=2\mu _{i}{\bf n}_{i}$) in the magnetic field ${\bf H}$%
. Slonczewski\cite{21} showed that carriers tunneling via the planar
junction mediate an exchange interaction between FM electrodes, which proves
to become of AFM type for sufficiently high barrier. As in that case we will
assume each two grains $i,j$ to be coupled by an AFM exchange $%
J_{ij}=J_{0}\exp (-2\kappa d_{ij})$, with $J_{0}<0$. So the energy will take
the form 
\begin{equation}
E=-\sum_{i<j}{\bf n}_{i}\cdot {\bf J}_{ij}\cdot {\bf n}_{j}+2\sum_{i}\mu _{i}%
{\bf n}_{i}\cdot {\bf H}  \label{cmenergy}
\end{equation}
where a tensor ${\bf J}_{ij}$ includes both the dipolar and the above
exchange interactions. At $T\rightarrow 0$, $H\rightarrow 0$ a system with
the energy given by Eq.(\ref{cmenergy}) will tend to magnetically frustrated
state. Thus, to make AFM correlation between grains consistent with FM
ground state, we have to attribute the moments contributing to $R_{ij}$ (Eq.(%
\ref{rfactor}) ) to only a small part of the grains, most probably to GB.
Indeed, our magnetization measurements on LPMO show that at lowered
temperature the magnetization curve bends only slightly below the expected
saturation (see Fig.3).

In the subsequent discussion a simplified scheme is used. In accordance with
Refs.\cite{4,5,7} in which: (i) the exact conductance is replaced by an
averaged inverse of Eq.(\ref{rfactor}) and (ii) the averaging over the
spatial and magnetic variables is carried out independently. This gives for
the resistivity 
\begin{equation}
\rho (T,H)=\frac{\rho _{U}}{1+\varepsilon \langle \cos \vartheta
_{ij}\rangle }  \label{avresist}
\end{equation}
where $\rho_{U}\propto \langle r_{ij}\rangle $. Due to Eq.(\ref
{rfactor}) the factor $\rho _{U}$ is not expected to vary essentially at low 
$T$ and $H$; in this range the field dependence of the resistivity results
from that of $\langle \cos \vartheta _{ij}\rangle $. Thus, irrespective of
the type of frustrated state at $H=0$, Eq.(\ref{avresist}) describes the
cause of the low-field negative MR as being the rotation of the 'partial'
grain moments to a unique direction along the magnetic field.\cite{14} For
AFM correlation and sufficiently low $H$, $\langle \cos \vartheta
_{ij}\rangle $ increases upon heating from a negative value at the ground
state (note that authors of Ref.\cite{7} estimated $\langle \cos \vartheta
_{ij}\rangle $ from their data to equal $\sim -0.8788$) to higher values.
Therefore, the part of Eq.(\ref{avresist}) depending on $\langle \cos
\vartheta _{ij}\rangle $ is a decreasing function of $T$. At $T=0$ such a
behavior persists until $\langle \cos \vartheta _{ij}\rangle $ changes its
sign at $H=H_{{\rm cr}}$, while at $H>H_{{\rm cr}}$ it becomes an increasing
function of $T$. To describe full dependence of $\rho (T,H)$ the modeling of 
$\rho _{U}$ is required. Zhang et al.\cite{4,5} obtained $U$ as a function
of the magnetization of the grain core and GB. Due to this model $\rho _{U}$
is an increasing function of temperature at $T<T_{\max }$.\cite{4,5} Thus,
in view of the above discussion concerning $\langle \cos \vartheta
_{ij}\rangle $, Eq.(\ref{avresist}) predicts that a minimum of $\rho (T,H)$
versus temperature should occur at a low $T$ and $H<H_{{\rm cr}}$. This
minimum will degrade upon increasing $H$ and disappear at $H\geq H_{{\rm cr}%
} $, in qualitative agreement with our data shown in Fig. 2, for which $H_{%
{\rm cr}}\approx 1.5$ T.

The quantitative use of Eq.(\ref{avresist}) is possible only under some
approximation to analytical calculation of $\langle \cos \vartheta
_{ij}\rangle $ and $\rho _{U}$ (otherwise, many-body simulation method may
be applied directly to the original network, so Eq.(\ref{avresist}) becomes
redundant). To illustrate the model we approximate $\langle \cos \vartheta
_{ij}\rangle $ by its value for a cluster of two grains with an AFM exchange 
$J$ and equal moments $\mu _{i}=\mu $. In addition, since the temperature
dependence of the in-grain and GB magnetization is dominated by the spin
wave $T^{3/2}$ terms at low well $T$, we assume 
\begin{equation}
\rho _{U}=r_{0}+r_{1}T^{3/2}  \label{swprefac}
\end{equation}
where $r_{0}$ and $r_{1}$ are the parameters independent on $H$. The model
parameters are then defined from the requirement that in the case $H=0$ Eq.(%
\ref{avresist}) fits the experimental data for $\rho (T,H)$ in the range
from $4.3$ K to $50$ K. In this fit Eq.(\ref{swprefac}) and the expression 
\cite{22} 
\begin{equation}
\langle \cos \vartheta _{ij}\rangle =-L\left( \left| J\right| /k_{{\rm B}%
}T\right) ,\;H=0,  \label{corrzh}
\end{equation}
where $L(x)=\coth (x)-1/x$ is the Langevin function, are taken into account.
This gives $\varepsilon =0.487$ $(P\approx 0.697)$, $\left| J\right| /k_{%
{\rm B}}$ $=155$ K, $r_{0}=0.145$ $\Omega \cdot {\rm cm}$, $%
r_{1}=1.026\times 10^{-4}\Omega \cdot {\rm cm}\cdot ${\rm K}$^{-3/2}$. We
used these parameters and a closed expression for $\langle \cos \vartheta
_{ij}\rangle $ at $H{{}\neq \,}0$ \cite{22} to calculate the model curves $%
\rho (T,H)$ at the increased values of the parameter $S=2\mu H/\left|
J\right| $ (see Fig.4).
\vskip .5cm 
\begin{figure}
\epsfxsize=3truein
\centerline{\epsffile{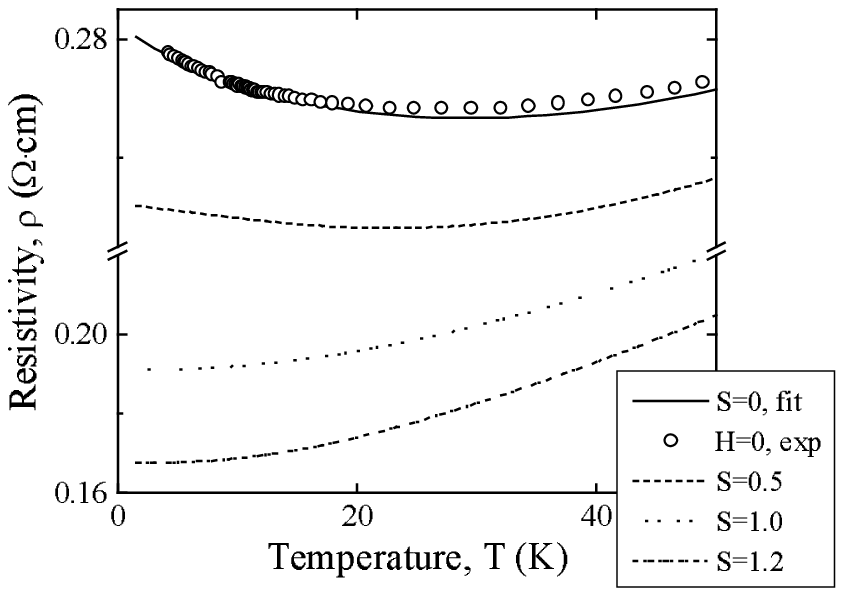}}
\label{Fig.4}
\end{figure}
%\vskip -11cm
\noindent
{\footnotesize {\bf FIG. 4.} 
Resistivity vs. temperature. The lines represent theoretical
curves based on the model of intergrain tunneling for various S (normalized
field). The open circles are the experimental results observed for La$_{0.5}$%
Pb$_{0.5}$MnO$_{3}$: 10 at. \% Ag.}
\vskip .5cm

It appears, that in the considered version of the model a quantitative 
agreement with the experiment cannot be
achieved with any choice of $\mu$ because of lacking the property of highest
magnetic-field response at the lowest fields observed in the experiment.
Really, the largest drop of $\rho (T,H)$ occurs experimentally at $0<H\leq
0.2$ T and at higher $H$ the change is slower (see Fig.2). The calculated
curves demonstrate much more gradual change in low-field region (see Fig.4).
Nevertheless, the predictions of the model concerning the generic behavior
of the resistivity minimum under the increase of $H$ are in a well
qualitative agreement with the experiment. Finally, if we equate the value
of $S\sim 1$ at which the minimum disappears in the model to its value at $%
H_{{\rm cr}}\approx 1.5$ T we retrieve $\mu \sim 200\mu _{{\rm B}}$. Low
values of $P$ and $\mu $ may express the fact that really a small part of a
grain with reduced magnetic order contributes to the tunneling.

\section{CONCLUSIONS}

A characteristic shallow minimum of the resistivity was found to occur in
various polycrystalline doped manganites La$_{1-x}$A$_{x}$MnO$_{3}$ where A
is Ca, Sr, Ba \cite{6,9,10,11,12} or Pb. \cite{13} Usually this minimum
occurs at low temperatures ($T<50$ K), shifts towards lower $T$ upon
applying a magnetic field and disappears at some field $H_{{\rm cr}}$. Two
models were considered to account for this minimum: (i) bulk scattering with
quantum corrections to conductivity and (ii) tunneling between AFM coupled
grains. The resistivity minimum in the first model, in disagreement with
experiment, is almost insensitive to $H$ in the range of interest. The
second model provides a fairly well qualitative description of the effects
observed. Even in a rough approximation it agrees with experiment.

Very recent experimental data also strongly support the suggested nature of
the effect considered. For example, the minimum of $\rho (T,0)$ and its
gradual vanishing under external $H$ are observed in a ceramic sample of La$%
_{0.8}$Sr$_{0.2}$MnO$_{3}$ while no such an effect is detected on a single
crystal of the same composition.\cite{6} The same effect is observed in an
epitaxial bicrystal film of La$_{0.67}$Ca$_{0.33}$MnO$_{3}$ with an
artificially created single grain boundary, but is absent in an epitaxial
film of the same composition.\cite{17} It also should be noted that the
spin-dependent tunneling of charge carriers in ceramic manganites discussed
in this paper is similar to that observed earlier in granular ferromagnetic
media - see, for example. \cite{23,24}

\section{ACKNOWLEGMENTS}

This research was supported by the Israeli Science Foundation administered
by the Israel Academy of Sciences and Humanities. The authors thank to Dr.
A. Shames for the help in our experiments.

%\end{multicols}

\end{document}